\documentclass[%
prx,
reprint,
superscriptaddress,
nofootinbib,
 amsmath,amssymb,
 aps,
floatfix,
]{revtex4-2}

\usepackage{graphicx}
\graphicspath{
{./img/}
}

\usepackage[caption=false]{subfig}

\usepackage{lineno}
\modulolinenumbers[5]

\let\oldalign\align
\let\oldendalign\endalign
\renewenvironment{align}{%
    \linenomathNonumbers\oldalign%
    }{%
    \oldendalign\endlinenomath%
    }

\let\oldequation\equation
\let\oldendequation\endequation
\renewenvironment{equation}{%
    \linenomathNonumbers\oldequation
    }{%
    \oldendequation\endlinenomath%
    }

\usepackage{graphicx}
\usepackage{bm}

\usepackage[dvipsnames]{xcolor}
\usepackage{scrextend}
\usepackage{makecell}
\usepackage{relsize}

\usepackage{booktabs}

\begin{document}

\title{Synthetic graphs for link prediction benchmarking}

\author{Alexey Vlaskin}
\email{avla7439@uni.sydney.edu.au}
 \affiliation{%
 School of Mathematics and Statistics, University of Sydney, New South Wales 2006, Australia
 }%
\author{Eduardo G. Altmann}%
\email{eduardo.altmann@sydney.edu.au}%
 \affiliation{%
 School of Mathematics and Statistics, University of Sydney, New South Wales 2006, Australia
 }%

\date{\today}

\begin{abstract}
Predicting missing links in complex networks requires algorithms that are able to explore statistical regularities in the existing data. Here we investigate the interplay between algorithm efficiency and network structures through the introduction of suitably-designed synthetic graphs. 
 We propose a family of random graphs that incorporates both micro-scale motifs and meso-scale communities, two ubiquitous structures in complex networks. A key contribution is the derivation of theoretical upper bounds for link prediction performance in our synthetic graphs, allowing us to estimate the predictability of the task and obtain an improved assessment of the performance of any method. Our results on the performance of classical methods (e.g., Stochastic Block Models, Node2Vec, GraphSage) show that the performance of all methods correlate with the theoretical predictability, that no single method is universally superior, and that each of the methods exploit different characteristics known to exist in large classes of networks. Our findings underline the need for careful consideration of graph structure when selecting a link prediction method and emphasize the value of comparing performance against synthetic benchmarks.
We provide open-source code for generating these synthetic graphs, enabling further research on link prediction methods.

\end{abstract}

\maketitle


\section{Introduction}

The problem of link prediction consists in using existing graph information (nodes and links) to predict unseen or missing links. Link prediction is an important and widely studied research topic in different disciplines -- including complex networks~\cite{liben-nowell_link_2003,clauset_hierarchical_2008,guimera_missing_2009,adamic} and machine learning~\cite{prauc,gnnSurvey,shopping} -- due to its connection to a variety of applications, e.g., predicting friendships among social-network users~\cite{social}, shopping recommendations~\cite{shopping}, predicting topics in large sets of documents~\cite{topics}, predicting protein interactions~\cite{proteins, TCP2019} and political views~\cite{polictical}, and facilitating scientific discoveries by linking previously disconnected disciplines~\cite{autoscience}.  

The number of different methods for link prediction has grown tremendously in the last decades, covering a variety of different approaches. The classification of such methods in link prediction surveys ~\cite{adamic,prauc,gnnSurvey,social,kumarsurvey,survey} include:
\begin{itemize}
    \item {\bf Similarity based methods}, which use local information of common neighbours of nodes and similarity measures (e.g., Jaccard or Adamic-Adar indices) for link prediction~\cite{adamic,TCP2019,L2L3exp}.
    \item {\bf Probabilistic models}, which combine statistical inference and random graph models (e.g., Stochastic Block Models~\cite{SBM,guimera_missing_2009,SBM-peixoto,graph-tool_2014,valles-catala2018}) to predict the probability of unobserved links. 
    \item {\bf Embedding based methods,} which use deep learning techniques to acquire node embeddings and a classifier (trained on the observed links) to predict links. Methods for constructing the embeddings typically go beyond the properties of simple nodes and include:
    \begin{itemize}
   \item {\bf Walk based methods}, which use random walks within the graph (e.g., Node2Vec ~\cite{node2vec} and DeepWalk\cite{deepwalk}).  
    
    \item {\bf Graph neural network methods (GNNs),} which go beyond immediate neighbours of nodes and explore larger neighbourhoods (e.g., GraphSage ~\cite{GraphSage} and SEAL ~\cite{SEAL}). 
    \end{itemize}
    
\end{itemize}

The evaluation of the different methods is typically based on the comparison of their performance in empirical networks (see, e.g.~\cite{opengraph,L2L3exp}). 
The limitations of this approach is that the performance depends on the specific networks used in the test and the reasons for the success or failure of the methods remain unclear. In the related problem of community-detection, these two limitations have been addressed  through the use of synthetic generated graphs~\cite{grphbenchmark}.  A further motivation for considering synthetic graphs to evaluate link-prediction methods, as we do in this paper, comes from Ref.~\cite{expander}, which reports that barbell graphs are especially challenging for GNNs.

In this paper, we incorporate the view that there is no best method for all graphs (in line with "no free lunch theorems"~\cite{nflt,peel_2017}) and that a primary goal of link-prediction research is to identify the extent into which different methods capture different network connectivity structures, with focus on structures that are known to generically exist in complex networks. For instance, some methods may be best for picking up micro-scale motifs (e.g., closing triangles~\cite{adamic,L2L3exp,TCP2019}), while others excel in detecting meso-scale structures~\cite{SBM,SBM-peixoto} (e.g., large communities of interconnected nodes). Following the approach used in Ref.~\cite{grphbenchmark} to evaluate {\it community-detection} methods, our goal here is to explore the method-structure connection in {\it link-prediction}
methods through the use of synthetic graphs.  We create synthetic graphs which have micro-scale motifs and meso-scale communities, two widespread properties of complex networks, but that at the same time are sufficiently simple to allow for the analytically computation of the ideal prediction score (i.e., the upper bound for the prediction achieved by any method). We then apply $4$ link-prediction methods to different parameter choices of our synthetic networks and find that our approach allows us to understand which methods are more suitable for different structures or networks. This indicates that our approach is not only successful in creating benchmarks to compare methods but it can also be used to better understand and improve specific methods.

\section{The link prediction problem}

\subsection{Problem definition}

We consider an undirected network (or graph) $G(V, E)$, where $V$ represents the set of nodes (vertices) -- which we can index $n=1, \ldots, N = |V|$ -- and E represents the set of links -- $\ell =1, \ldots |L|$ links (edges). We denote by $E_{O}$ the set of \textbf{observed links} and by $E_{U}$ the set of \textbf{unobserved links}  in the graph so that $E = E_{O} \cup E_{U}$ and $ E_{O} \cap E_{U} = \emptyset$. We also define the set of possible yet non-existing links as $\widetilde{E}$, such that $E \cap \widetilde{E} = \emptyset$ and $|\widetilde{E}|+|E| = N(N-1)/2)$. In this paper we consider the link prediction problem of finding the unobserved links in $E_{U}$  given the observed links $E_{O}$ and all nodes $V$. 

In practice, in most empirical networks $G(V,E)$, we do not have access to unobserved true links so that the evaluation of link-prediction methods is performed by dividing the $E$ links of an observed network randomly into two sets $E_{U}$ and $E_{O}$. In line with previous studies of link prediction~\cite{node2vec,prauc,SEAL}, here we adopt this strategy and choose $10\%$ of the links to be unobserved, i.e., $\frac{|E_{U}|}{|E|} = 0.1$. 

 \subsection{Quantification of the performance}\label{ssec.t}

We consider that the output of link-prediction methods is a set of predicted links $E_p \subseteq E_{U} \cup \widetilde{E}$. The size of this set, $0 \le |E_p| \le |E_{U} \cup \widetilde{E}|$, is typically controlled by a threshold $0\le t \le 1$ that sets the rate of link prediction (e.g., $E_p$ is composed by all links predicted with probability $p\ge t$ or by the top $t$ percentage of most-likely links). The number of true positives (TP) is given by $|E_p(t) \cap E_{U}|$, false positives (FP) by $|E_p(t) \cap \widetilde{E}|$, true negatives (TN) by $|\widetilde{E} \setminus E_p(t)|$, and false negatives (FN) by $|E_{U} \setminus E_p(t)|$. The true positive rate (TPR) and false positive rate (FPR) are then defined as
\begin{equation}\label{eq.tpr}
TPR(t) = \frac{TP(t)}{TP(t) + FN(t)},
\end{equation}
\begin{equation}\label{eq.fpr}
FPR(t) = \frac{FP(t)}{FP(t) + TN(t)}.
\end{equation}
In this paper we use as a measure of the quality of the link prediction the Area Under the Curve ($0\le AUC\le 1$) formed in the FPR(t) vs. TPR(t) graph when varying $t$ in $[0,1]$, defined as
\begin{equation}\label{eq.auc}
AUC = \int_{0}^{1} TPR(FPR^{-1}(t)) \,dt.
\end{equation}
We use the AUC measure because it is threshold free, constrained and normalized to $[0; 1]$(the larger the $AUC$, the better the prediction), and it has a well-defined value $AUC=0.5$ for a random prediction (null model). 

In the evaluation of the AUC, we sample an equal number of positive unobserved links and non-existing links, leading to a balanced number of positive and negative examples at the evaluation step, and use the same set of unobserved links and negative examples for all methods. This approach, aligned with the recommendations in Ref. ~\cite{graphevalnn}, is critical to ensure fairness in the comparison because existing links form a small subset of the total possible links and, otherwise, the quantification of the performance could be affected by the fact that there are much more negative samples than positives ones. In the results reported below, we construct ten graphs per configuration and show the average and variance of the methods' performance (AUC) over these graphs. 

\begin{figure*}[!bt]
    \centering
    \includegraphics[width=0.8\textwidth]{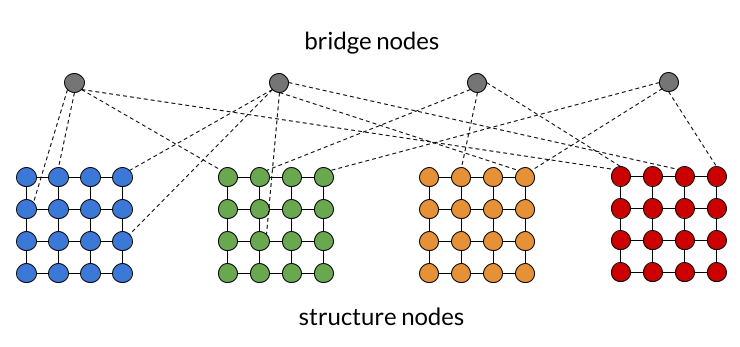}
    \caption{An example of the synthetic graphs we use as benchmarks for link-prediction methods. In this example, there are $N_{B}=4$ bridge nodes and $M=4$ structures. Each structure is a $k=4$ 2d lattice so that the number of structure nodes is $N_{S} = 4 \times 4 \times 4$. Bridge nodes connect randomly to structural nodes only with probability $D_{B} / N_{S}$, with $D_b=4$ used in this example.}
    \label{fig.illustration}
\end{figure*}
 
\subsection{Link-prediction methods} \label{sec.methods}

In this work we report results on four widely-used link prediction methods, each representing its own class:

\begin{itemize}
    \item The classical Adamic-Adar similarity coefficient, which predicts links based on the common neighbours of each node~\cite{adamic}. This method was introduces for link prediction on social networks, where triangles are known to be common (friends of my friend are my friends) too). 
    \item An inferential approach based on Stochastic Block Models (SBM)~\cite{SBM,guimera_missing_2009,valles-catala2018,SBM-peixoto}, in which the prediction of the probability of a (missing) link between two nodes depends exclusively on the block assignment of the nodes.  We use the Bayesian microcanonical formulation of SBM reviewed in Ref.~\cite{SBM-peixoto} and implemented in Ref.~\cite{graph-tool_2014}, where the number and partition of nodes in blocks (latent variables) are inferred based on the minimization of the Description Length measure (which accounts for both the likelihood and complexity of the model).
    \item Node2Vec~\cite{node2vec} family of methods conducts random walks inside the graph that are used to learn node embedding through a procedure called skip-gram. The model jointly trains a classifier which predicts the probability of the link given a learned node embedding of two nodes. We use Gemsim's implementation available in Ref.~\cite{rehurek2011gensim}.
    \item GraphSage~\cite{GraphSage} also uses random walks inside the graph and constructs a node embedding, but constructs a multi-layer graph neural network on top of the learned node embedding incorporating at the same time the observed graphs structure. We use Stellagraph's implementation described in Ref.~\cite{StellarGraph}.
\end{itemize}
We choose these methods because they are widely used, are representative of the four different approaches for link prediction methods discussed in the Introduction (based on Ref.~\cite{survey}), and have simple implementations that facilitate their interpretability, a crucial issue  to understand the connection between link-prediction methods and the structures present in networks. See our repository~\cite{REPO} for further details on the implementation of these methods.

\section{Synthetic graphs} \label{sec.randomgraphs}

In this section we introduce the family of random-graph models to be used as benchmarks to study link-prediction methods. 
In order to explore the dependence of the performance of the methods on the network structure, we propose a graph generating procedure which incorporates different generic properties of complex networks.  In particular, we are interested in incorporating micro-scale (e.g., triangles, motifs) and meso-scale (e.g., communities) structures within the same connected component of a sparse network. Varying the strength of each of these structures reveal the connection between the performance of different methods and the structures, i.e., the extent into which the methods can explore the regularities induced by these structures to successfully predict missing links. We keep the method for creation of synthetic graphs simple in order to obtain an analytical value for the performance of the best possible method. This is essential in order not only to compare methods against each other, but also to evaluate how far a method is from the maximum-possible performance and to evaluate in which extent variations of the performance are affected by the difficulty of the link-prediction task. 

\subsection{Generation procedure}

We consider that the graph has two types of nodes, structural (s) and bridge (b) nodes. The $N_s$ structural nodes are divided in $M$ different structures and there are no links between nodes of different structures. Within each structure, the connectivity pattern follows the same simple rule, e.g., all-to-all (clique), a 2d lattice, or a 2d lattice with closed diagonals. The $N_b$ bridge nodes connect only to structure nodes, at random with the same fixed link probability 
 $\frac{D_{B}}{N_{S}}$, where $D_b$ is the expected degree of bridge nodes (a parameter of the model). Figure~\ref{fig.illustration} illustrates one possible realization of these graphs with $N_b=4$ bridge nodes and $M=4$ structures ($2-d$ lattices).

More formally, we define an allocation function $\tau(i)$ that given the index $i \in [1,N]$  of a node returns the node type $\tau(i) \in \{1, \ldots, M; B\}, $ where $[1; M]$ corresponds to the index of one of the $M$ structures and $B$ means that the node is a bridge nodes.
 We also define a link function $\phi(i,j)$ that, for two nodes with indices $i,j$ in the same structure, returns 0, if a link does not exist for this structure and 1 if the link exist.  According to the generating procedure described above, the probability of a link between any nodes $i,j \in [1; N], i > j$ can be written as
%
%
where the delta function $\delta$ is defined as $\delta(x) = 1$ if and only if $x=0$ (otherwise, $\delta(x)=0)$, the if condition is satisfied only if the output of the operation with the delta functions is $1$, and we ensure that the adjacency matrix is symmetric $A(i,j)=A(j,i)$ (un-directed graph). Table~\ref{tab.parameters} lists the parameters of our model together with derived quantities of interest. An algorithmic description of the generation process is in~\ref{app.genproc} and the implementation in our repository~\cite{REPO}.

\begin{table*}[!bt]
\begin{center}
\begin{tabular}{ |c|c|c| } 
 \multicolumn{3}{c}{\bf Parameters of the synthetic graph} \\
 \hline
  $N_{B}$ & Number of bridge nodes & \\
   \hline
  $D_{B}$ & Average bridge degree &   \\
   \hline
  M & Number of structures &  \\
    \hline
 Structure ($\tau$)  & Type of structure  & (e.g., 2-d lattice, clique, etc.) \\
 k & size of each structure &  \\   
  \hline
  \multicolumn{3}{c}{} \\
  \multicolumn{3}{c}{\bf Derived quantities} \\
  \hline
      $N_{S}$ & Number of structure nodes &  $N_{S} = M k^2$ (for 2-d lattice) \\
  \hline 
   N & Number of nodes in the graph & $N = N_{S} + N_{B}$ \\
 \hline
   $C_{S}$ & Ratio between $N_{S}$ and $N$ & $C_{S} = N_{S} / N $ \\
    \hline
    \multicolumn{3}{c}{} \\
  \multicolumn{3}{c}{\bf Expected number of edges} \\
    \hline
       $E_{S}$ & Number of edges within each structure & $E_{S} = \sum_{\delta(\tau(i)=\tau(j))} \phi(i,j)$ \\
     \hline
       $E_{B}$ & Number of bridge edges in the graph & $E_{B} = D_{B} \cdot N_{B}$\\
       \hline
         $E$ & Number of edges in the graph & $E = D_{B} \cdot N_{B} + M \cdot E_{S}$\\
    \hline
$\overline{e_{SS}}$ & \begin{tabular}{@{}c@{}} Fraction of all existing links \\ that are between structure nodes \end{tabular}& $\frac{ME_{S}}{E}$\\
\hline
$\widetilde{e_{SB}}$ & \begin{tabular}{@{}c@{}}Fraction of all non-existing links \\  that are between structure and bridge nodes \end{tabular} & $\frac{N_{B} \cdot (N_{S} - D_{B})}{\frac{N(N-1)}{2}-E}$\\
\hline
\end{tabular}
\caption{List of parameters and main quantities of our synthetic random-graph model. See~\ref{app.edges} for other expressions for the number of (non-existing) edges and their relationship to the type of structure structure.}
\label{tab.parameters}
\end{center}
\end{table*}

\subsection{Predictability of synthetic graphs}\label{ssec.predictability}

Here we quantify the predictability of missing links in the synthetic graph by computing the maximum possible AUC score. This is done considering an {\it ideal} algorithm that\footnote{We recall that the unobserved (missing) links $E_{U}$ in our graph are created by removing $10\%$ of the links after the generation of the graph and our assumption is that this does not prevent us from identifying the structure $[1,M]$ of nodes.}: 

\begin{itemize}
    \item[(i)] predicts all structural links perfectly;
    \item[(ii)] predicts all bridge links with the same fixed probability ($0<(D_{B}/N_{S})<1$);
    \item[(iii)] attributes a zero probability for the links that, by construction, can not possibly exist (e.g., between bridge nodes and between nodes of different structures).
\end{itemize}

\begin{figure}
    \includegraphics[width=0.8\linewidth]{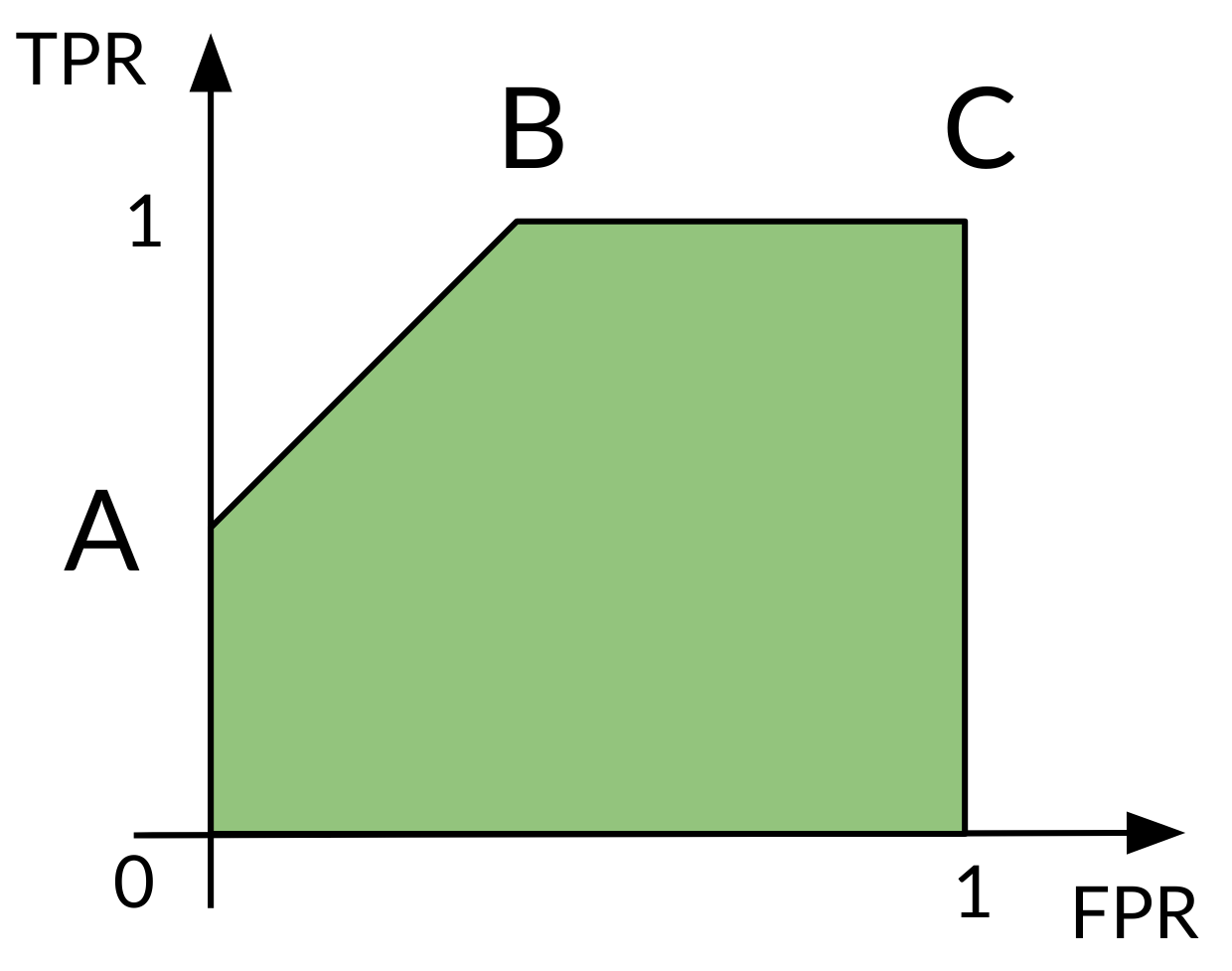}
    \centering
    \caption{Maximum prediction score in the synthetic graphs. The straight lines $\vec{AB}$ and $\vec{BC}$ compose the prediction curve $(FPR(t),TPR(t))$ for $t\in[0,1]$ of the ideal prediction method in our synthetic graph. The point A is located at $(0,\widetilde{e_{SB}})$, the point $B$ at $(\overline{e_{SS}},1)$, and C at $(1,1)$. See~Tab.~\ref{tab.parameters} for the formulas expressing $\widetilde{e_{SB}}$ and $\overline{e_{SS}}$ as a function of model parameters. See~\ref{app.analyticalideal} for a derivation of these points.}
    \label{fig.idealdots}%
\end{figure}

 The importance of this {\it ideal} algorithm is that its AUC is an upper bound for the performance of any link-prediction method. To compute this value,  we show in Fig.~\ref{fig.idealdots} the TPR(t) vs FPR(t) curve of the ideal algorithm. Its key properties can be understood as follows:
\begin{itemize}
\item The possibility of predicting exactly the missing structure links -- point (i) above -- implies that the ideal method can achieve a $TPR > 0$ at $FPR=0$. The maximum value of the TPR with FPR=0 (point A in Fig.~\ref{fig.idealdots}) is given by the fraction of all existing links that are between structure nodes, denoted by $\overline{e_{SS}}$.
\item The knowledge of the links that are not possible to exist -- point (iii) above -- implies that the ideal method can predict all missing links ($TPR=1$) at a $FPR<1$. The minimum value of $FPR$ with $TPR=1$ (point B in Fig.~\ref{fig.idealdots}) is given by the fraction of all non-existing links that are between structures and bridge nodes, denoted by $\widetilde{e_{SB}}$.
\item Between these regimes, the curve (connecting A and B in Fig.~\ref{fig.idealdots}) is a straight line because of the fixed probability $D_{B}/N_{S}$ of the bridge-structure links -- point (ii) above. 
\end{itemize}
The AUC is then computed geometrically from Fig.~\ref{fig.idealdots}  -- the full area of the square (1) minus the area of the top left triangle connecting A, B, and (1,1) -- as
\begin{equation}\label{eq.ideal}
AUC = 1 - \frac{1}{2}(\widetilde{e_{SB}} \cdot (1 - \overline{e_{SS}})).
\end{equation}
We see that there are two factors that can increase the predictability (AUC) of link-prediction tasks in our synthetic graphs:
\begin{itemize}
    \item  increasing the fraction of predictable links within each structure (i.e., increasing $\overline{e_{SS}}$). These links will typically be associated to micro-scale motifs placed within the structures.
    \item decreasing  the fraction of non-predictable non-links between structure and bridge nodes (i.e., decreasing $\widetilde{e_{SB}}$). This happens because the bridge-structure links (existing and non-existing) are the only ones that are uncertain to be predicted in our model (ideal method). 
\end{itemize}
The values of $\overline{e_{SS}}$ and $\widetilde{e_{SB}}$ can be expressed explicitly as a function of the parameters of our synthetic network, as shown in~\ref{app.analyticalideal}. 

Similar computations as the ones shown above to the ideal method can be used to obtain the AUC for the SBM method with a block allocation equal to $\tau(i)$ (i.e., one block for each of the $M$ structures and one block for the bridge nodes $B$). Such {\it planted SBM} will make an ideal prediction of the bridge-structure links, an ideal prediction that there are no bridge-bridge links, and will also correctly predict no link between nodes of different structures. The crucial difference to the ideal prediction method is that the planted SBM will predict a fixed probability for any link within the same structure, while the ideal method will predict such links precisely. The derivation of the AUC for the planted SBM is given in~\ref{app.analytical}.

\section{Numerical Results}
\begin{figure}
    \centering   
    \includegraphics[width=1\linewidth]{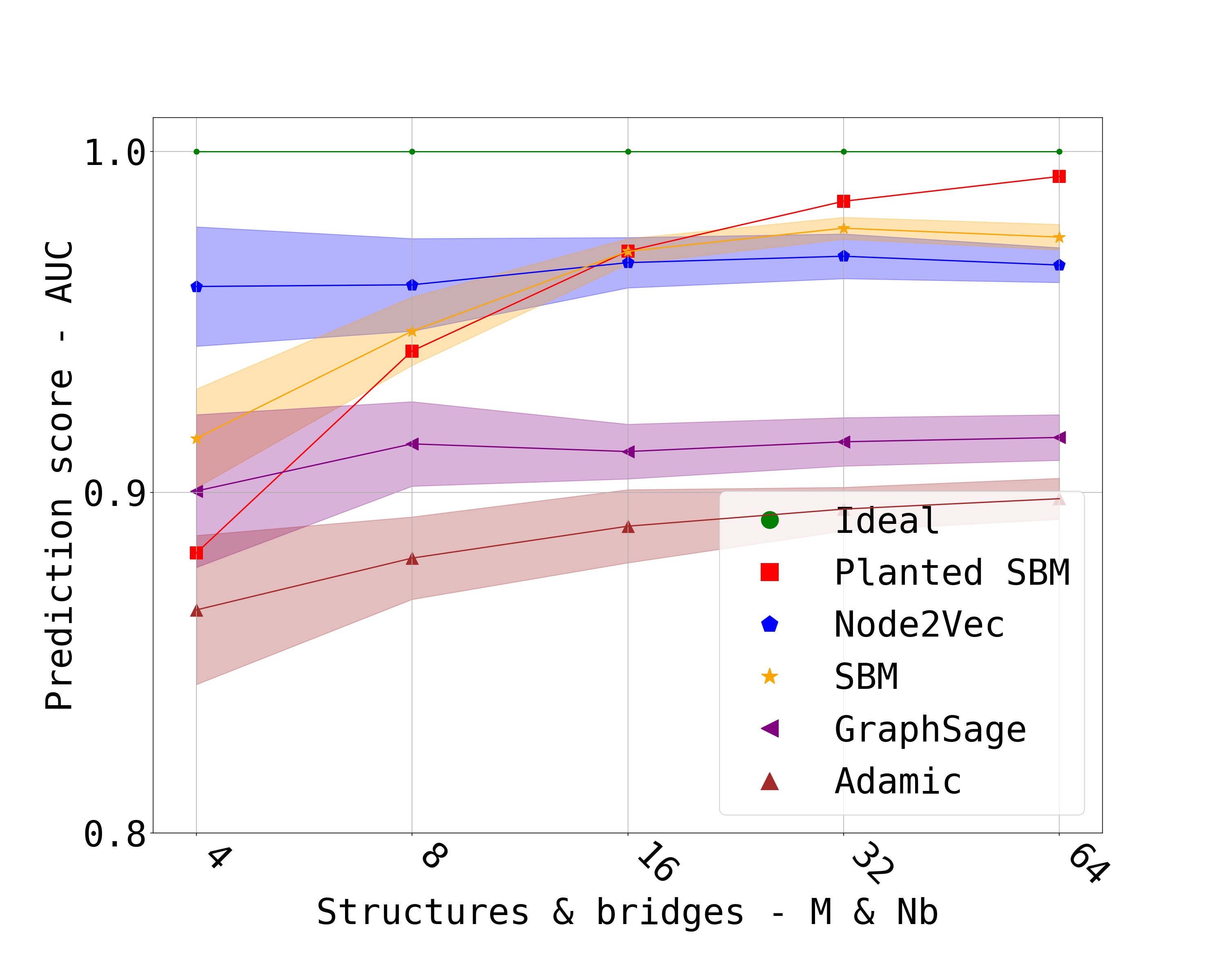}
    \caption{Performance of link prediction methods (see legend) in networks with increasing number of structures $M$ and bridge nodes $N_b$.  
    The networks have average bridge degree $D_{B} = 5$,  the structures are 2-d lattices with $k=8$ and closed diagonals, and various $N_{B} = M$  (x-axis). 
    }
    \label{fig.latticediag}\label{fig.3}%
\end{figure}
In this section we describe the numerical results obtained by applying the four link prediction methods (described in Sec.~\ref{sec.methods}) to the synthetic graphs (generated as described in Sec.~\ref{sec.randomgraphs}), comparing the numerical outcomes to the theoretically maximum score computed in Eq.~(\ref{eq.ideal}). 
\begin{center}
  
\begin{figure}
    \includegraphics[width=0.95\linewidth]{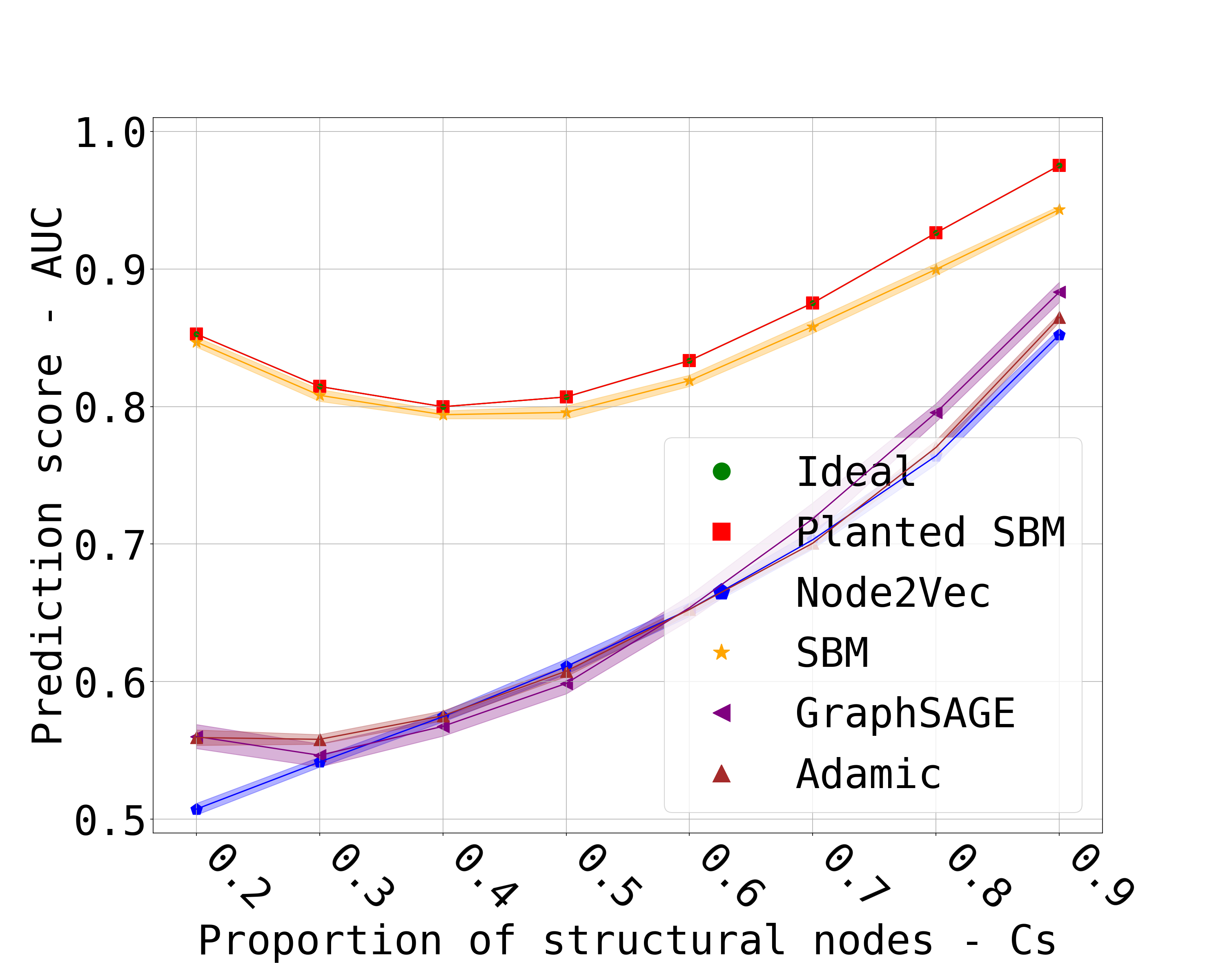}
    \includegraphics[width=0.95\linewidth]{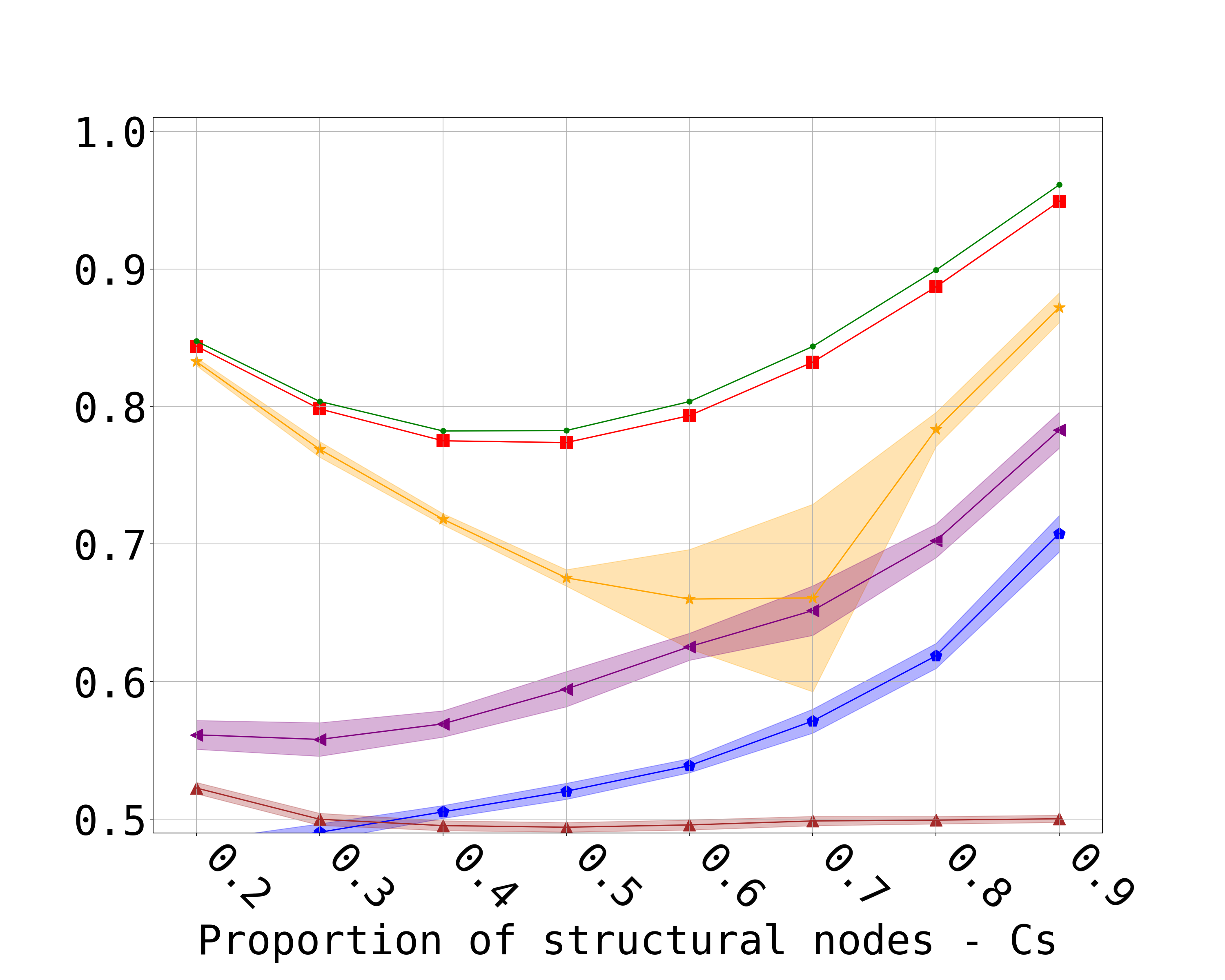}
    \caption{Performance of link prediction methods for graphs with increasing fraction $C_{S}$ of structure nodes. Graphs with $N = 3200$ nodes were build using two types of structures: cliques (left panel) and 2d lattices (right) and bridges. $k = 8$, so clique has 8 nodes each and lattice has $8 \times 8 = 64$ nodes each. Probability of bridge  $\alpha = 12 / ( N \cdot C_{S} )$. Please see ~\ref{app.tables} - ~\ref{faaa-table-clique} and ~\ref{faab-table-lattice} for the exact generation parameters.}
    \label{fig.latticevscliques}\label{fig.4}
\end{figure}
\end{center}
\subsection{Micro-scale motifs vs. meso-scale structures}\label{ssec.micromeso}

We start our numerical investigations by considering the extent into which link-prediction methods exploit two ubiquituous properties of complex networks: micro-scale motifs (triangles, squares, etc.) and meso-scale structures (communities, blocks of densely connected nodes, etc.). To address this question, we consider a scenario of our synthetic graphs in which the number $M$ of structures equals the number of bridge nodes $N_b$ (as in Fig.~\ref{fig.illustration}). Increasing $N_b=M$, keeping the type and size $k$ of each structure constant, we increase the overall relevance of the meso-scale structures (distinct structures $M$) in the network maintaining constant the over-all presence of motifs (inside each of the $M$ structures). 

Figure~\ref{fig.latticediag} shows the numerical outcome for one particular realization of the scenario described above.  It shows that the performance of both the Node2Vec and GraphSage methods remain roughly constant, confirming that these methods rely mostly on micro-scale motiffs. In contrast, the performance of the SBM method grows significantly with $M=N_b$, confirming the connection between meso-scale structures and blocks in SBM (the difference between the numerical results in SBM and the theoretical computation in the planted SBM will be discussed in Sec.~\ref{ssec.sbms} below). For $N_b=M \rightarrow \infty$, the AUC of SBM approaches the ideal case. The constant high value of the ideal prediction is explained because the choice of $k=8$ (i.e., $8$ by $8$ lattice structures) creates graphs rich on meso-scale structures and micro scale motifs where random links are in minority, implying that the number of (predictable) structural links $E_s$ is much larger than the (unpredictable) bridge links $E_b$. More importantly, the numerical results show that Node2Vec outperforms SBM when only a small number of structures are present (small $M=N_b$), but the opposite is true in the opposite limit (large $M=N_b$). This illustrates the central points motivating our investigations: that there is no one link-prediction method that performs best in all cases and that their performance is directly connected to different (predictable) structures in the network.

\subsection{Varying micro-scale structures}

In the example above we considered a large network  $N\rightarrow \infty$ scenario in which we deliberately kept constant both the maximum predictability (ideal case) and the strength of micro-scale motifs. In this section we we vary these two aspects and explore how they affect the performance of the different link-prediction methods.

We first consider a case in which balances between graphs with dominating micro-structures and graphs where random links prevail. To explore that, we vary the ratio $C_{S}$ between the number of bridge and structural nodes by constructing synthetic graphs with various numbers of structures $M$ and bridge nodes $N_b$ while keeping $N$ fixed. Intuitively the more structural nodes we have in the graph (i.e., the higher the $C_{S}$) the better for methods that capture the micro structure like Node2Vec and GraphSage. 
This expectation is confirmed numerically in Figure \ref{fig.latticevscliques}
for two realizations of this scenario (two differente structure types). Besides the direct relationship between the performance of Node2Vec and GraphSage with $C_{S}$, these numerical results show that the performance of SBM follows the same non-monotonic dependence with $C_{S}$ observed in the ideal case. This reveals that this behavior is not a property alone of SBM, but of its ability to explore the predictable information in the graph. More generally, this point shows the importance of our computation of the theoretical value for the ideal prediction, and our introduction of synthetic graphs that allow for such computation. Finally, comparing the two scenarios -- full clique structures (left) and lattice (right)- we see that GraphSage is noticeably better than Node2vec in the sparser lattice case, while on clique graphs GraphSage and Node2Vec have about the same performance. GraphSage use of clique nodes neighbours does not provide benefits, while on lattices GraphSage  provides a better structural representation. The simple Adamic-Adar method shows a similar behaviour, with a performance comparable to Node2Vec and GraphSage in the case of clique structures (left panel, when all triangles are closed) but a performance indistinguishable from random guessing (AUC=0.5) in the case of lattice structures (right panel, no triangles). We also confirm that as randomness decreases (increase in $C_S$), the the performance of the structural methods (Node2Vec and GraphSage) increase as expected.
SBM can describe the clique case perfectly -- AUC of planted SBM equal to the ideal case on the left -- but not the lattice case -- AUC for SBM smaller than ideal case on the right.

\begin{figure}[!bt]
    \includegraphics[width=0.95\linewidth]{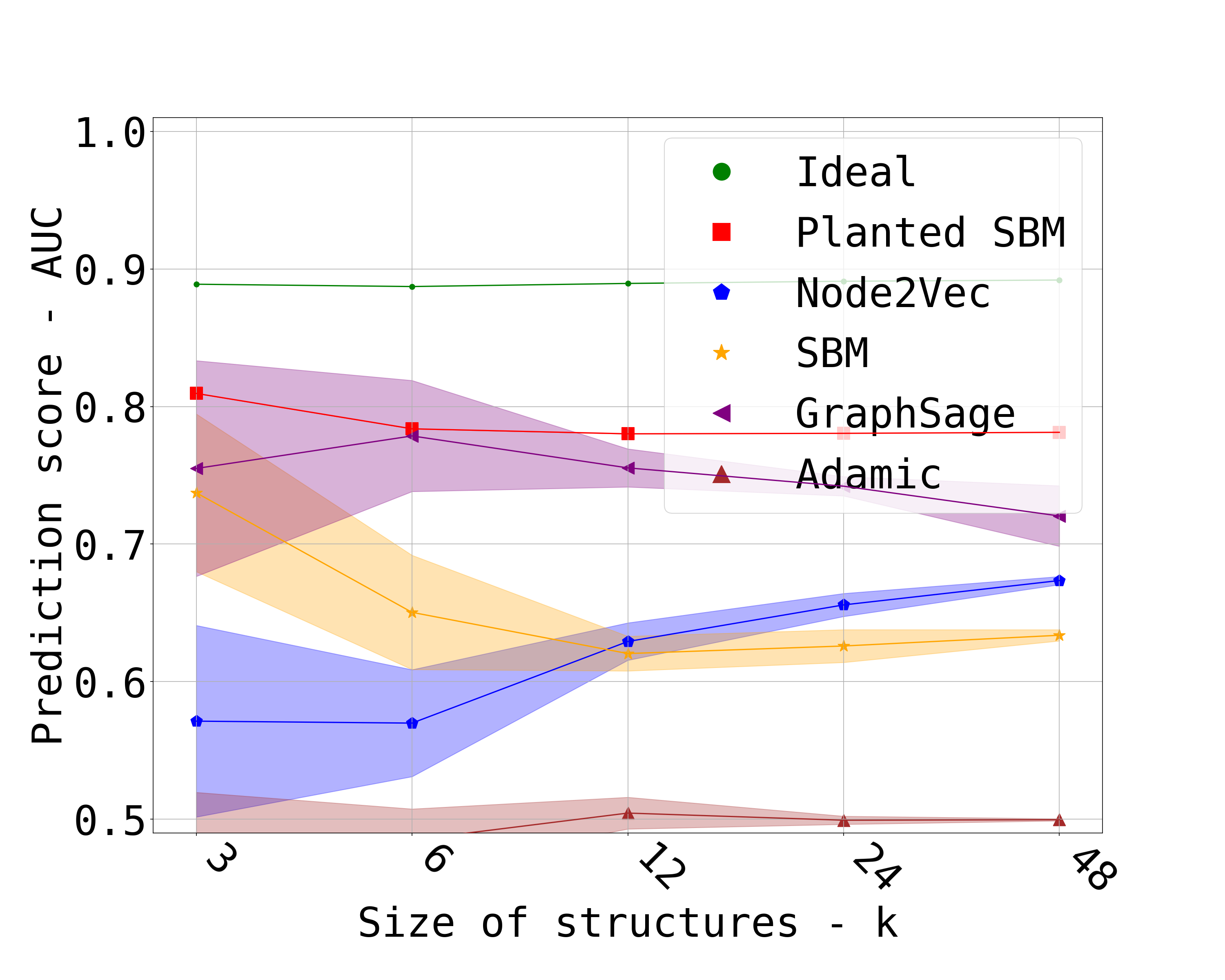}
    \includegraphics[width=0.95\linewidth]{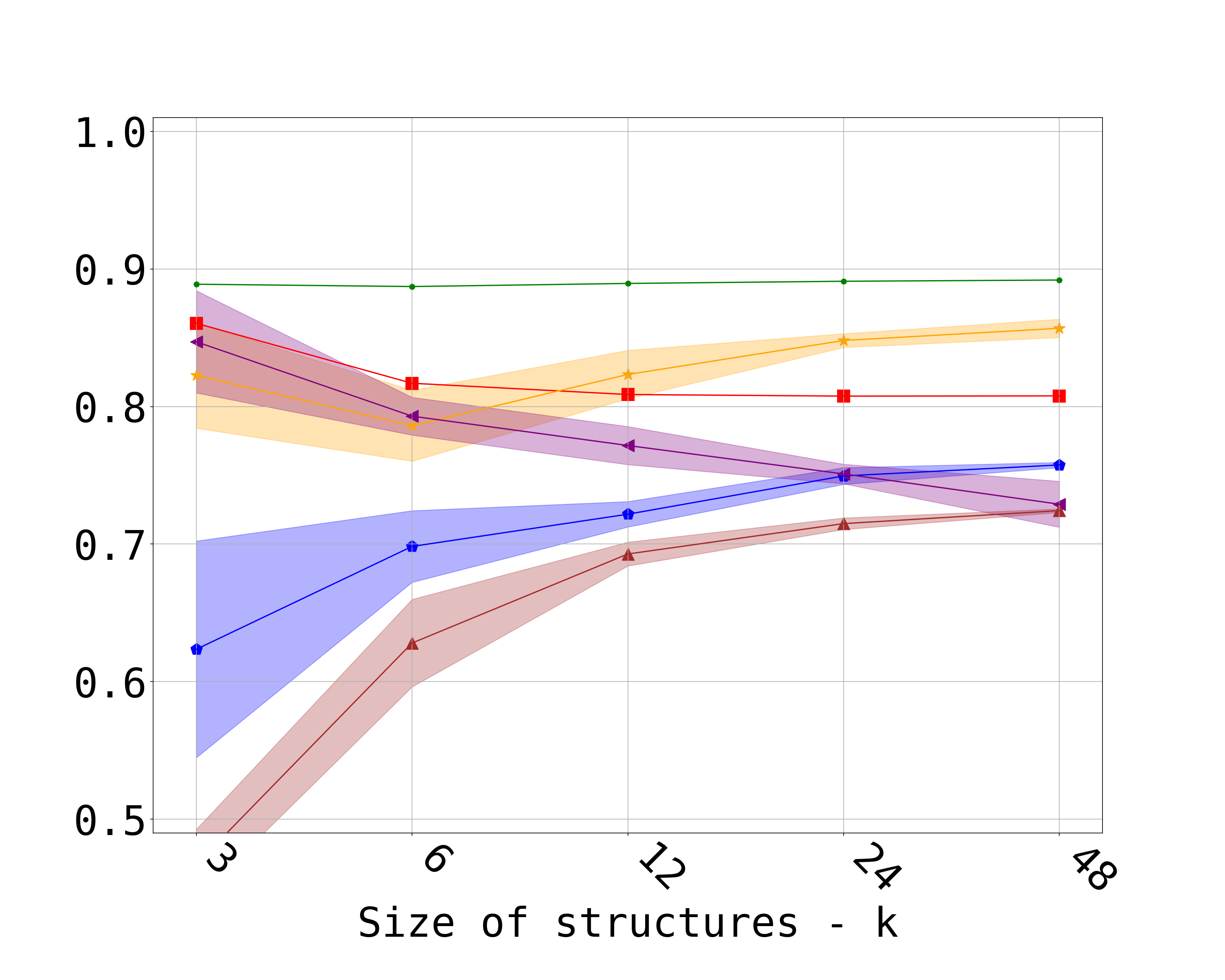}
    \caption{Performance of link prediction methods for structures of growing size $k$.  The number of structures $M = 4$, the ratio of structure nodes $C_{S} = 0.75$, and the average bridge degree $D_{B}$ are kept fixed. Left: we use square 2d lattices, where diagonals are not closed. Right: we use square 2d lattices with closed one diagonal forming closed triangles, as illustrated in Fig.~\ref{fig.latticegraphwithdiags}. In both cases, the number of structure nodes grows with $k$ as $N_s = 4k^2$ and bridge nodes as $N_{B} = (1 - 0.75)\cdot N$, to keep $C_{S}=0.75$ constant.
    }
    \label{fig.latticegnnbest}\label{fig.5}%
\end{figure}

\begin{figure}[!ht]
    \centering
    \includegraphics[width=0.95\linewidth]{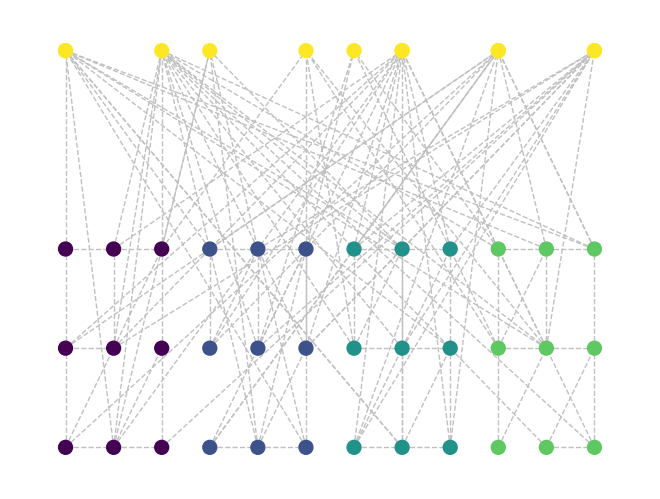}
    \caption{One realization of the synthetic graph used for link prediction in Fig.~\ref{fig.latticegnnbest} (right, $k=3$). The graph has bridge nodes (yellow cluster) and $M=4$ structures formed by square 2d lattices with closed diagonals. In this graph, 10\% of links were removed as used in the evaluation of link-prediction methods.}
    \label{fig.latticegraphwithdiags}%
\end{figure}

The second scenario we consider is to increase the size $k$ of structures, keeping the ratio $C_{S}$ between structure and bridge nodes constant. The large network limit $N\rightarrow \infty$ is achieved increasing $k\rightarrow \infty$ but keeping the number of structures $M$ fixed. Figure \ref{fig.latticegnnbest} shows the results for $C_{S} = 0.75$  -- balanced between structures and randomness -- and two types of structures -- 2-d lattice (left) and 2-d lattice with closed diagonals (right). 
One example of the graph we used -- 2-d lattice with closed diagonals -- is shown in Fig.~\ref{fig.latticegraphwithdiags}.
Growing the size $k$ corresponds to a growing importance of micro-scale motifs, which in this case consists of triangles (as in the case considered on the right of Fig.~\ref{fig.latticegnnbest}) and squares (in both left and right cases of Fig.~\ref{fig.latticegnnbest}). In both cases the predictability of the link-predication problem (ideal case) is less than one and insensitive to $k$. This means that, in these cases, variations on the performance of the different methods reflect their ability to explore the micro-scale motifs and not the difficulty of the problem as a whole. Looking at the trend of these methods with $k$ leads to interesting conclusions:
\begin{itemize}
\item[(i)] The Adamic-Adar method shows the expected sensitivity to the existence of closed triangles (increase in performance in the right plot) and the expected lack of sensitive to the existence of squares (no change in performance in the left plot, indistinguishable from random guessing). These results are also consistent with the recent findings~\cite{TCP2019,L2L3exp} that limitations of 2-hop-based similarity measures (such as the Adamic-Adar method) can be overcome by considering 3-hop-based similarity measures (which can detect squares). 
\item[(ii)] In both cases (with triangles and squares), Node2Vec increases its performance with $k$ while GraphSage decreases, with Node2Vec getting better than GraphSage for large $k$. One possible explanation for GraphSage's decline is that our choice of fixed training parameters (e.g. number of epoch, batch size) and model configuration (e.g. number of layers, size of the internal representation) might be less effective for the larger graphs obtained for large $k$\footnote{A fine tuning of parameters for growing graphs is possible, but would increase considerably the computational cost.}
\end{itemize}
Altogether, the results show that different regimes in $k$ have GraphSage, Node2Vec, and SBM as the best method, corroborating the idea that no single method is the best for all regimes. 

\subsection{SBM vs. planted SBM}\label{ssec.sbms}

\begin{figure}[!bt]
\includegraphics[width=0.95\linewidth]{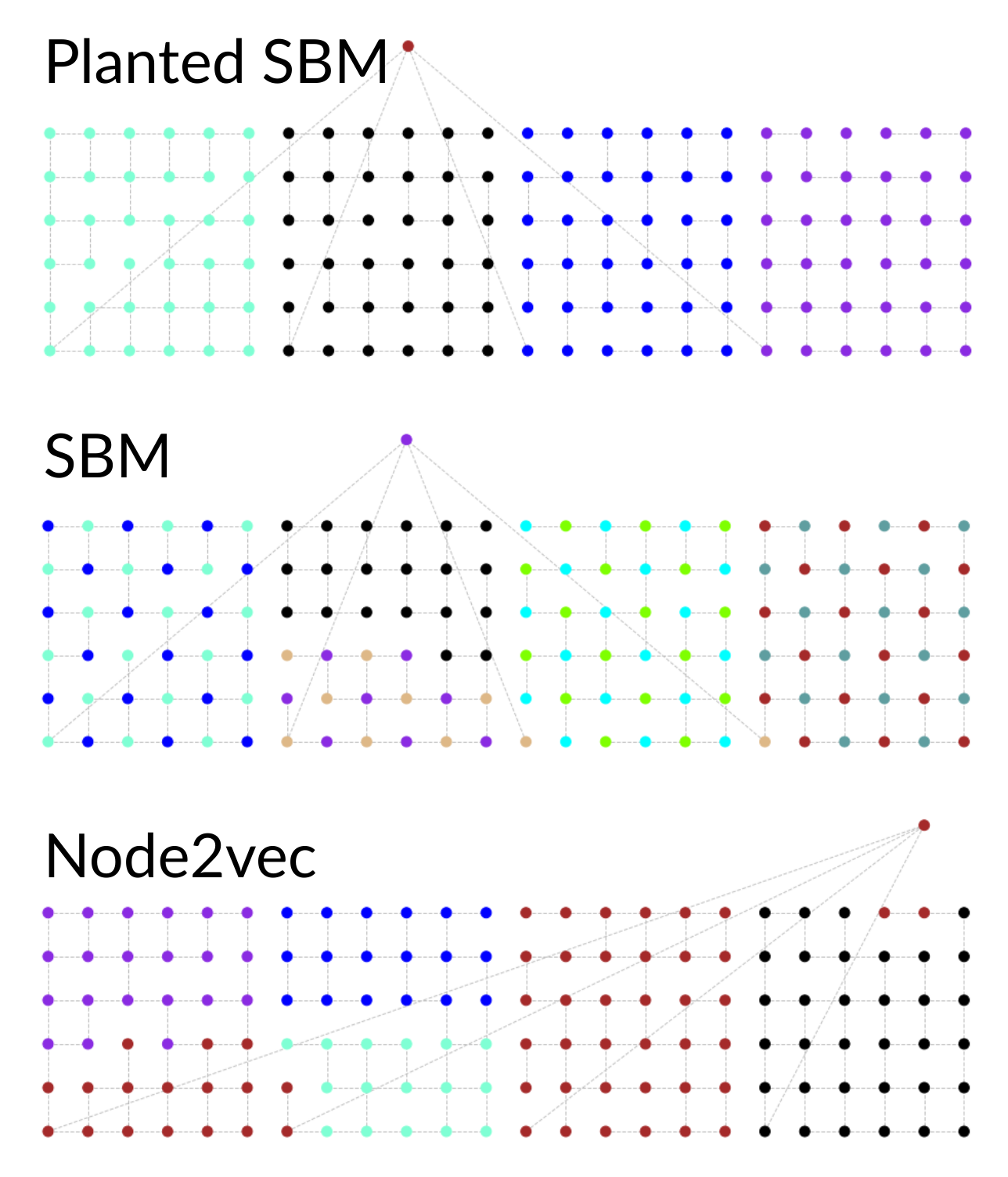}
    \centering
    \caption{Group allocation of nodes (different colours) obtained from different methods. From top to bottom, the panels show the group allocations according to the planted SBM, the numerical SBM (inferred), and Node2Vec (clustering of nodes in node embedding). In this example, the inferred partition of SBM (middle panel) outperforms the planted SBM (top panel) in link prediction. In all cases, the network used for link prediction is shown after the removal of $10\%$ of the links. It corresponds to a realization of our synthetic graphs with $M=4$ structure, a 2d lattices with $k=6$ as structure type, and a single $N_b=1$ bridge node with $D_b=4$. }
    \label{fig.plantedvssbm}\label{fig.7}%
\end{figure}

A recurring observation in the figures reported in the previous two sections is that the performance (AUC) of the SBM method  $AUC_{SBM}$ can be equal, smaller, and larger than the analytically computed planted SBM result $AUC_{Planted SBM}$. To understand why this happens -- the goal of this section -- we need to consider that the planted partition is one of the possible partitions of the numerical implemented SBM method we consider (namely, the one in which there are $M+1$ blocks, one for each structure of the synthetic graph and one for the bridge nodes). This planted partition is a natural partition of the graph that can, in principle, be discovered during the SBM inference (which explores possible block allocation of the nodes). This is the reason that curves of SBM and planted SBM largely follow each other  $AUC_{SBM} \approx AUC_{Planted SBM}$, a numerical observation which reassures our theoretical understanding. Assuming that the choice of block allocation (model selection) aligns with a high link prediction scores, the typical case as discussed in Ref.~\cite{valles-catala2018}, this implies that $AUC_{SBM}\ge AUC_{Planted SBM}$. We note also that our synthetic graphs are within the SBM class only for specific types of structures (e.g., cliques) and therefore, in general, we do not have a theoretical guarantee that $AUC_{SBM} \le AUC_{Planted SBM}$.

The reasoning above suggests that the observations $AUC_{SBM} < AUC_{Planted SBM}$ are due to the inability of the inference process we use for SBM to find the planted partition. We confirmed that this is the case by considering one network used in Figure~\ref{fig.latticegnnbest} and computing the preference of the SBM method (description length) of the planted partition (21478.5) and of the inferred partition (21560.8). The preference for the planted partition (smaller description length) confirms that the planted partition is considered better by SBM, which was unable to find it during the inference procedure. While this suggests the need for considering better inference methods, here we stick to out-of-the-shelf algorithms in order to obtain a comparison across different methods using standard parameter choices (as used in the study of empirical networks, when no planted partition is available). 

To understand the case that $AUC_{SBM} > AUC_{Planted SBM}$, we considered parameters in which this was observed and compared the SBM partition obtained numerically with the planted partition. One example is shown in Figure \ref{fig.plantedvssbm}, obtained when the structure corresponds to a 2-d lattice (i.e., beyond the SBM generative process). We see how SBM achieves a better prediction score by splitting the nodes within each structure in two blocks, with links exclusively across them. This shows that the high performance of SBM is based on its flexibility to accommodate additional patterns within structures. For comparison, we clustered nodes in groups using the embeddings created in Node2Vec and we find that they align also in a great extent with the allocation in groups in structures, as done in the planted SBM.

\section{Conclusions}

In summary, we introduced a family of random graphs to be used as synthetic benchmarks for link-prediction algorithms. We generated these graphs  with a controllable mix of micro-scale motifs and meso-scale communities, allowing for the exploration of the dependence of different algorithms on these properties. Importantly, these graphs allow for a closed form computation of the performance of an ideal algorithm, which is an upper bound for the performance of any algorithm and can be seen as a quantification of the predictability of the link-prediction problem. The importance of this is that it allows us to better assess the performance of algorithms, not only by measuring how far they are from the ideal case but also to disentangle effects affecting the variation in performance from the overall difficulty of the link-prediction task. For instance, scenarios in which the performance of algorithms follows the predictability revealed by the ideal one -- such as the non-monotonic dependence of the SBM method in Fig.~\ref{fig.4} -- reveal that the variations in the performance of the method are due to the variation of the difficulty of the problem under consideration. In contrast, scenarios in which the maximum performance remains constant but the performance of algorithms change -- such as in Figs.~\ref{fig.3} and \ref{fig.5} -- reveal the connection between the algorithms and specific structures being varied in the synthetic graphs (e.g., micro-scale motifs, meso-scale structures.

Our numerical findings for different networks and four link-prediction algorithms (Adamic-Adar, SBM, Node2Vec and GraphSage) confirm that there is no one method that performs best in all graphs~\cite{nflt,peel_2017,L2L3exp}. For example, SBM captures the meso-scale arrangement of communities~(Figs.~\ref{fig.3} and~\ref{fig.4}) but can struggles with micro-scale motifs within communities (Fig.~\ref{fig.5}), while Node2Vec and GraphSage show the opposite tendency. We found that GraphSage seems to perform better than Node2Vec on more difficult problems (smaller structures, larger mixture of random and deterministic links as in Figs.~\ref{fig.4} and~\ref{fig.5}), but Node2Vec beats GraphSage when structures have a lot of micro-scale motifs (Fig.~\ref{fig.3}). We also found that SBM shows a great flexibility that allows it to increase the number of blocks and outperform the expected (planted) SBM when the underlying graph is far from an SBM generative process (e.g., structures as 2d lattices with large $k$, as shown in Figs.~\ref{fig.5} and~\ref{fig.7}). 

Our results illustrate the potential of using synthetic graphs to reveal the connection between network structures and performance of link-prediction algorithms. We hope that our results and openly-available codes~\cite{REPO} will motivate the proposal of other families of random graph models to be used for this purpose, which could capture other types of structures known to exist in complex networks (e.g., skewed degree- and structure-size- distributions~\cite{grphbenchmark}, motifs in directed graphs, and hierarchies~\cite{clauset_hierarchical_2008}). In turn, this could lead to the development of improved link-prediction methods through the combination of existing methods and thus the exploration of different structures and statistical regularities. 

\section*{Acknowledgements}
We thank Lamiae Azizi for discussions during the design of this project.


%

\appendix
\clearpage

\begin{widetext}
  \section*{Appendices}
\section{Generating procedure}\label{app.genproc}

The graph-generation algorithm implemented in our repository~\cite{REPO} can be summarized as:

\begin{itemize}  
\item[1.] Select the primary parameters: 
\begin{itemize}
    \item $N_{B}$ - number of bridge nodes in the graph.
\item $D_{B}$ - average Degree of bridge nodes. 
\item $M$ - number of structures in the graph.
\item $k$ - number of nodes on the side of a square lattice. 
\end{itemize}

\item[2.] Divide the $N_S= M k^2$ structure nodes in the $\tau=1, \ldots, M$ structures, choose the structure type from [Clique, Lattice, Lattice with diagonals], and connect the nodes inside the same structure as:
\begin{itemize}
    \item Clique: fully connected sub-graph, i.e.,  $\phi(i, j) = 1$ for all $i, j$ in the same structure $\tau(i)=\tau(j)$, leading to a total of $ k \cdot (k-1) / 2$ links. 
\item Lattice: a $2d$ square lattice, i.e., $\phi(i, j) = \delta(|i - j| = k) \cdot \delta(|i - j| = 1)$, with indexes $i,j$ enumerated sequentially, leading to $2 \cdot k \cdot (k -1)$ links.
\item Lattice With Diagonals: as in the $2d$ square lattice defined above, but with $m=1$ or $m=2$ diagonals inside the lattice closed, leading to $2 \cdot k \cdot (k -1) + m \cdot k^2$ links.
\end{itemize}
\item[3.] Generate $N_{B}$ bridge nodes and connect each of them to each of the $N_S$ structure nodes with a probability $D_B/N_S$. 
\end{itemize}

\section{Notation for edges in generated graph}\label{app.edges}

Here we write the expression for the number of existing $\overline{E}$, possible but non-existing~$\widetilde{E}$, and potential $E$ links as a function of the model parameters (see also Tab.~\ref{tab.parameters} in the main text). In general, for each class $X$ defined below, $E_{X} = \widetilde{E_{X}} + \overline{E_{X}}$. The different sets of links are illustrated in Fig~\ref{fig.appendix} and can be expressed as follows:

\begin{itemize}
\item $E_{SS}$: number of all possible links between structured nodes, that are within structure blocks. 
\begin{equation}
E_{SS} = \sum_1^M \omega(k) \cdot (\omega(k) - 1) / 2 = \sum_{i=1}^{N} \sum_{j=1+1}^{N} \delta(\tau(i) == \tau(j))    
\end{equation}
\item $\overline{E_{SS}} = M \cdot E_{S}$: number of existing links between structured nodes, that are withing structure blocks. Summed over all structures and depends on the concrete structure type. 
\begin{equation}
\widetilde{E_{SS}} = E_{SS} - \overline{E_{SS}}
\end{equation}
\item $\widetilde{E_{SS}}$: number of possible, but not existing links between structured nodes within the same structure. 
\item $E_{SB}$: Set of all possible links between bridges and structured node. 
\begin{equation}
E_{SB} = N_{S} \cdot N_{B} = N \cdot c \cdot N \cdot (1 - c) 
\end{equation}
\item $\overline{E_{SB}} \equiv E_{B}$: Number of existing links between bridges and structured nodes.  
\begin{equation}
\overline{E_{SB}} = D_{B} \cdot N \cdot (1 - c) = \sum_{i=1}^{N} \sum_{j=i+1}^{N} \delta((\tau(i) \neq \tau(j)) \cdot (\tau(j) = B))
\end{equation}
\item $\widetilde{E_{SB}}$: number of possible, but not existing links between bridges and structured nodes. 
\begin{equation}
\widetilde{E_{SB}} = E_{SB} - \overline{E_{SB}} = N_{S} \cdot N_{B} - \overline{E_{SB}}
\end{equation}
\item $\widetilde{E_{BL}}$: number of possible, but not existing links between different structured blocks. 
\begin{equation}
\widetilde{E_{BL}} = \frac{N_{S} \cdot (N_{S} - 1)}{2} = \frac{N \cdot c \cdot (N \cdot c - 1)}{2}
\end{equation}
\item $\overline{E_{BL}}$: number of existing links between different structured nodes. Empty set.
\begin{equation}
\overline{E_{BL}} = 0
\end{equation}
\item $\widetilde{E_{BB}}$: number of possible, but not existing links between bridge nodes. 
\begin{equation}
\widetilde{E_{BB}} =  \frac{N_{B} \cdot N_{B}}{2}=  \frac{N^{2} \cdot (1 - c)^{2}}{2}
\end{equation}
\item $\widetilde{E} = \widetilde{E_{BL}} + \widetilde{E_{BB}}$: number of possible, but not existing links outside of blocks of structures and bridges.
\end{itemize} 

An exact expression for the cases that depend on the structure type is give in Tab.~\ref{tab.cases}.

\begin{table*}[!bt]
\begin{center}
\begin{tabular}{ |c|c|c|c| } 
 \hline
 Group & Cliques & Lattice & Lattice with diagonals \\ [1ex] 
  \hline
 $E_{SS}$  &  $ \frac{N\cdot c \cdot(k-1)}{2}$ & $ \frac{2\cdot N \cdot c \cdot (k - 1)}{k}$ & $\frac{2\cdot N \cdot c \cdot (k - 1)}{k}$ \\ 
  \hline
 $\overline{E_{SS}}$ & $ \frac{N \cdot c \cdot (k-1)}{2}$ & $ 2 \cdot k \cdot (k - 1) \cdot \frac{N \cdot c}{k^2}$ &  $ (2 \cdot k \cdot (k - 1) + (k - 1)^2) \cdot \frac{N \cdot c}{k^2}$ \\
  \hline
\end{tabular}
\caption{Number of (possible) links between structures for the different lattice types (Cliques, 2d lattice, and 2d lattice with diagonals).}\label{tab.cases}
\end{center}
\end{table*}

\begin{figure}
    \centering
    \includegraphics[width=0.5\textwidth]{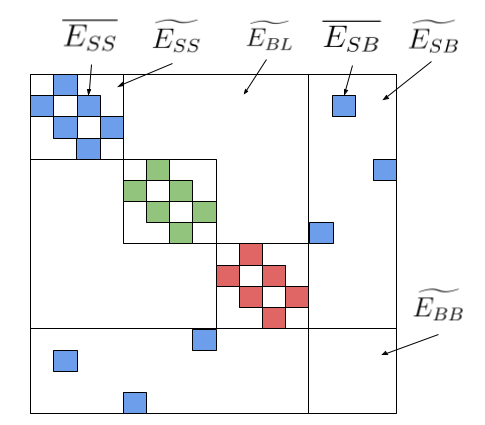}
    \caption{Graph adjacency matrix with different sets of edges indicated. Coloured squares represent connected vertices.}\label{fig.appendix}
\end{figure}

\section{Performance of the ideal algorithm}\label{app.analyticalideal}

Here we elaborate on the computation of the AUC for the ideal algorithm performed in Sec.~\ref{ssec.predictability}, as a function of the parameters of our synthetic graphs. We consider the dependence of the true-positive rate (TPR) and false-positive rate (FPR) on the probability threshold $0 \le t \le 1$ (starting from which a link is predicted) defined in Sec.~\ref{ssec.t}. The quantification of successful and unsuccessful predictions for all $t$ is shown in Tab.~\ref{tab.ideal}. These results allow us to calculate the AUC from Eq.~(\ref{eq.auc}) score from the AUC definition in Eq.~(\ref{eq.auc}) -- or, equivalently, geometrically from Fig.~\ref{fig.idealdots} -- as
\[
AUC = 1 - 0.5 \cdot ( 1 - \frac{\overline{E_{SS}}}{\overline{E_{SS}}+\overline{E_{SB}}}) \cdot \frac{\widetilde{E_{SB}}}{\widetilde{E_{SB}}+\widetilde{E}}
\]
Defining  $\widetilde{e_{SB}}= \frac{\widetilde{E_{SB}}}{\widetilde{E_{SB}}+\widetilde{E}}$ and $\overline{e_{SS}} =  \frac{\overline{E_{SS}}}{\overline{E_{SS}}+\overline{E_{SB}}}$  we retrieve Eq.~(\ref{eq.ideal}) derived in the main text.

\begin{table*}
\begin{center}
\begin{tabular}{ |c|c|c|c|c|c|c|c| } 
 \hline
 Point & Threshold & TP & FN & FP & TN & TPR & FPR \\ 
 \hline
  C & $t = 0$ & $\overline{E_{SS}} + \overline{E_{SB}}$ & 0      & $\overline{E_{SS}}+\widetilde{E}$& 0 & 1 & 1 \\
  \hline
  B & $0 < t \le P$ & $\overline{E_{SS}} + \overline{E_{SB}}$ & 0 & $\widetilde{E_{SB}}$     & $\widetilde{E}$ & 1 & $\frac{\widetilde{E_{SB}}}{\widetilde{E_{SB}} + \widetilde{E}}$ \\ 
  \hline
  A & $P < t \le 1$ & $\overline{E_{SS}}$ & $\overline{E_{SB}}$     & $0$       & $\widetilde{E_{SB}}+\widetilde{E}$ & $\frac{\overline{E_{SS}}}{\overline{E_{SS}}+\overline{E_{SB}}}$ & 0 \\ 
 \hline
\end{tabular}
\caption{Performance of the prediction of the ideal method for different thresholds $t$. The definition of the prediction rates (true positive TP, true negative TN, false positive FP, and true negative TN) and threshold $t$ are given in Sec.~\ref{ssec.t}, with the true- and false positive rates (TPR and FPR) expressed in Eqs~(\ref{eq.tpr}) and~(\ref{eq.fpr}), respectively. The ideal method is presented in Sec.~\ref{ssec.predictability}. The number of edges $E$ are expressed as a function of the model parameters in~\ref{app.edges} The points A,B, and C denoted in the first column are illustrated in Fig.~\ref{fig.idealdots}.}\label{tab.ideal}
\end{center}
\end{table*}

\section{Analytical calculations for planted SBM}\label{app.analytical}

\begin{figure*}[!bt]
    \centering
    \includegraphics[width=0.8\textwidth]{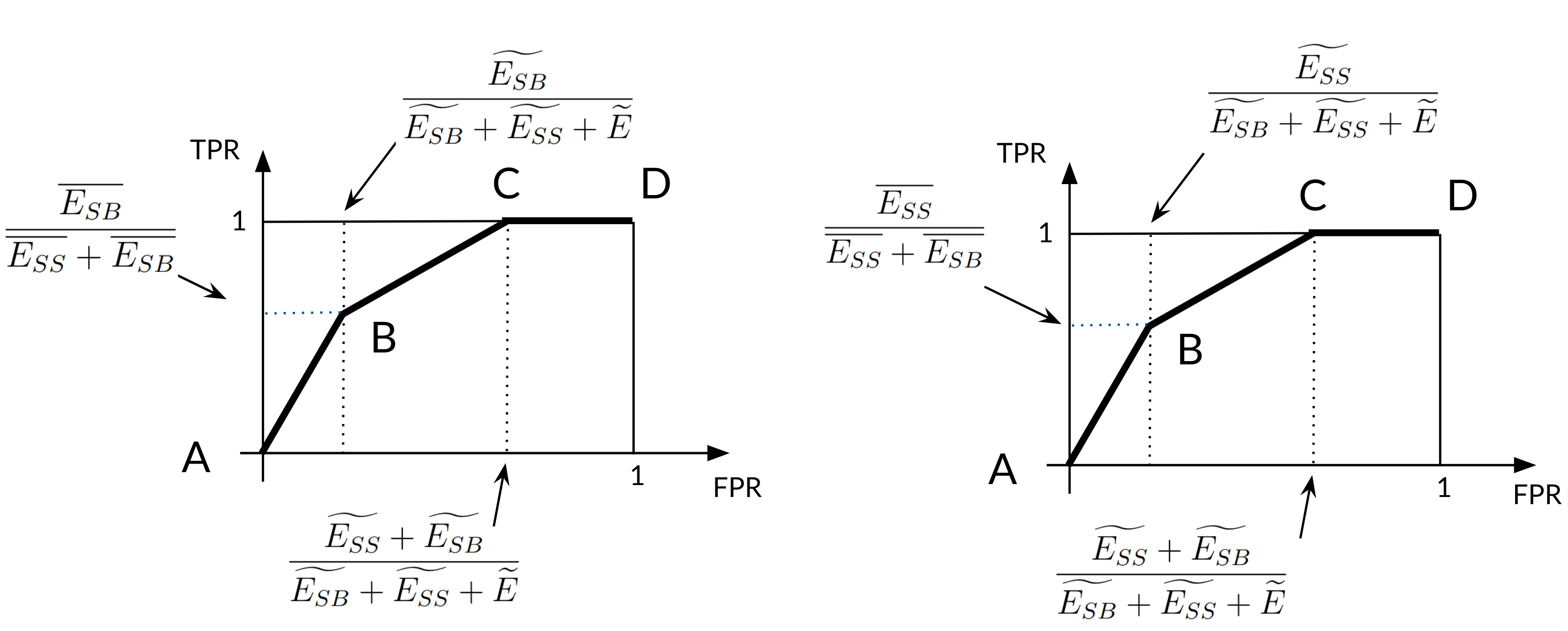}
    \caption{Computing the AUC for the planted SBM. Left $p>q$, right $q>p$.}\label{fig.planted}
\end{figure*}

\begin{table*}
\tiny
\begin{center}
\begin{tabular}{ |c|c|c|c|c|c|c|c| } 
 \hline
 Points & Threshold & TP & FN & FP & TN & TPR & FPR \\ 
 \hline
  D & $t = 0$ & $\overline{E_{SS}} + \overline{E_{SB}}$ & 0  & $\widetilde{E_{SS}} + \widetilde{E_{SB}} + \widetilde{E}$ & 0 & 1 & 1 \\
  \hline
  C & $ t < q$ & $\overline{E_{SS}} + \overline{E_{SB}}$ & 0 & $\widetilde{E_{SS}} + \widetilde{E_{SB}}$     & $\widetilde{E} $ & 1 & $\frac{\widetilde{E_{SS}} + \widetilde{E_{SB}}}{\widetilde{E_{SB}} + \widetilde{E_{SS}} +\widetilde{E}}$ \\ 
  \hline
  B & $q \le t<p$ & $\overline{E_{SB}}$ & $\overline{E_{SS}}$  & $\widetilde{E_{SB}}$ & $\widetilde{E_{SS}}+\widetilde{E}$ & $\frac{\overline{E_{SS}}}{\overline{E_{SS}}+\overline{E_{SB}}}$ & $\frac{\widetilde{E_{SB}}}{\widetilde{E_{SB}} + \widetilde{E_{SS}} +\widetilde{E}}$  \\ 
  \hline
  A & $p \le t \le 1$ & 0 & $\overline{E_{SS}} + \overline{E_{SB}}$ & $0$ & {\tiny $\widetilde{E}+\widetilde{E_{SB}}+\widetilde{E_{SS}}$} & 0 & 0 \\ 
 \hline
\end{tabular}
\caption{Threshold and TPR / FPR calculations for planted SBM for $p>q$. }\label{tab.planted1}
\end{center}
\end{table*}

\begin{table*}
\tiny
\begin{center}
\begin{tabular}{ |c|c|c|c|c|c|c|c| } 
 \hline
 Points & Threshold & TP & FN & FP & TN & TPR & FPR \\ 
 \hline
  D & $t = 0$ & $\overline{E_{SS}} + \overline{E_{SB}}$ & 0  & $\widetilde{E_{SS}}+\widetilde{E_{SB}}+\widetilde{E}$ & 0 & 1 & 1 \\
  \hline
  C & $ t \le q$ & $\overline{E_{SS}} + \overline{E_{SB}}$ & 0 & $\widetilde{E_{SS}} + \widetilde{E_{SB}}$ & $\widetilde{E}$ & 1 & $\frac{\widetilde{E_{SS}} + \widetilde{E_{SB}}}{\widetilde{E_{SS}} + \widetilde{E_{SB}} + \widetilde{E}}$ \\ 
  \hline
  B & $q < t \le p$ & $\overline{E_{SS}}$ & $\overline{E_{SB}}$  & $\widetilde{E_{SS}}$       & $\widetilde{E_{SB}}+\widetilde{E}$ & $\frac{\overline{E_{SS}}}{\overline{E_{SS}}+\overline{E_{SB}}}$ & $\frac{\widetilde{E_{SS}}}{\widetilde{E_{SS}}+\widetilde{E_{SB}}+\widetilde{E}}$ \\ 
  \hline
  A & $p < t \le 1$ & 0 & $\overline{E_{SS}} + \overline{E_{SB}}$ & 0 & $\widetilde{E}+\widetilde{E_{SB}}+\widetilde{E_{SS}}$ & 0 & 0 \\ 
 \hline
\end{tabular}
\caption{Threshold and TPR / FPR calculations for planted SBM for $q>p$. }\label{tab.planted2}
\end{center}
\end{table*}

Here we compute the performance (AUC) of the planted SBM method defined in Sec.~\ref{ssec.predictability} as a function of the parameters of our synthetic graphs. The computations are similar to the ones performed for the ideal method in~\ref{app.analyticalideal}. We denote by $q$ the constant SBM probability of a link within the same structure and by $p$ the probability of a link between a structure and bridge link. These probabilities can be obtained as the corresponding ratios between existing and all possible links as
\begin{equation}
p = \frac{D_{B}}{N_S}
\end{equation}
and
\begin{equation}
q = \frac{2 \cdot latticeLinks }{k^2(k^2-1)} = \frac{4(k-1)}{k \cdot (k^2-1)}.
\end{equation}
The AUC of the planted SBM depends crucially on whether $p>q$ or $p<q$ because this determines which links are predicted first (i.e., for smaller $t$). The results for the link-prediction performance at different prediction threshold $t$ is given, for each of these two cases, in Tabs.~\ref{tab.planted1} and~\ref{tab.planted2}. These values can be visualized in the TPR-FPR plot in Fig.~\ref{fig.planted}, which allow for the computation of the AUC in Eq.~(~\ref{eq.auc}) for $p>q$ as

\begin{equation}
AUC_{p>q} = 1 - \frac{1}{2} \cdot (1 - \frac{\overline{E_{SB}}}{\overline{E_{SS}} + \overline{E_{SB}}}) \cdot \frac{\widetilde{E_{SS}}}{\widetilde{E_{SB}} + \widetilde{E_{SS}} + \widetilde{E}}- \frac{1}{2} \cdot \frac{\overline{E_{SB}}}{\overline{E_{SS}} + \overline{E_{SB}}} \cdot \frac{\widetilde{E_{SB}}}{\widetilde{E_{SB}} + \widetilde{E_{SS}} + \widetilde{E}} - (1 - \frac{\overline{E_{SB}}}{\overline{E_{SS}} + \overline{E_{SB}}}) \cdot 
\frac{\widetilde{E_{SB}}}{\widetilde{E_{SB}} + \widetilde{E_{SS}} + \widetilde{E}},
\end{equation}

and, similarly, for $q > p$ as
\begin{equation}
  AUC_{q>p} = 1 - \frac{1}{2} \cdot (1 - \frac{\overline{E_{SS}}}{\overline{E_{SS}} + \overline{E_{SB}}}) \cdot \frac{\widetilde{E_{SB}}}{\widetilde{E_{SB}} + \widetilde{E_{SS}} + \widetilde{E}} -  \frac{1}{2} \cdot \frac{\overline{E_{SS}}}{\overline{E_{SS}} + \overline{E_{SB}}} \cdot \frac{\widetilde{E_{SS}}}{\widetilde{E_{SB}} + \widetilde{E_{SS}} + \widetilde{E}}  - (1 - \frac{\overline{E_{SS}}}{\overline{E_{SS}} + \overline{E_{SB}}}) \cdot \frac{\widetilde{E_{SS}}}{\widetilde{E_{SB}} + \widetilde{E_{SS}} + \widetilde{E}}.
  \end{equation}

\section{Exact parameter tables}\label{app.tables}
In Tabs.~\ref{faaa-table-clique} and~\ref{faab-table-lattice} we show the complete set of parameters used to construct some of the synthetic graphs used in our numerical results.

\begin{table}[!h]
\begin{center}
\begin{tabular}{ccccc|cccc}
\toprule
   $N_{B}$ &  M & $D_B$ &   $C_S$ &  Structure &  k & L &  $L_{B}$ &  $L / L_{B}$ \\
\midrule
640 &  80 & 12 & 0.2 &  Clique &  8 &   32960 &         30720 &   0.93 \\
960 &  120 & 12 & 0.3 &  Clique &  8 &   30240 &         26880 &   0.89 \\
1280 &  160 & 12 & 0.4 &  Clique &  8 &   27520 &         23040 &   0.84 \\
1600 &  200 & 12 & 0.5 &  Clique &  8 &   24800 &         19200 &   0.77 \\
1920 &  240 & 12 & 0.6 &  Clique &  8 &   22080 &         15360 &   0.70 \\
2240 &  280 & 12 & 0.7 &  Clique &  8 &   19360 &         11520 &   0.60 \\
2560 &  320 & 12 & 0.8 &  Clique &  8 &   16640 &          7680 &   0.46 \\
2880 &  360 & 12 & 0.9 &  Clique &  8 &   13920 &          3840 &   0.28 \\
\bottomrule
\end{tabular}
\caption{Generation parameters for set of graphs in Figure ~\ref{fig.latticevscliques} - Left.}\label{faaa-table-clique}
\end{center}
\end{table}

\begin{table}[!h]
\begin{center}
\begin{tabular}{ccccc|cccc}
\toprule
 $N_{B}$ &  $M$ & $D_B$) &   $C_S$ &  Structure &  k & L &  $L_{B}$ &  $L / L_{B}$ \\
\midrule
640 & 10 & 12 & 0.20 & Lattice & 8 & 31840 & 30720 & 0.96 \\
960 & 15 & 12 & 0.30 & Lattice & 8 & 28560 & 26880 & 0.94 \\
1280 & 20 & 12 & 0.40 & Lattice & 8 & 25280 & 23040 & 0.91 \\
1600 & 25 & 12 & 0.50 & Lattice & 8 & 22000 & 19200 & 0.87 \\
1920 & 30 & 12 & 0.60 & Lattice & 8 & 18720 & 15360 & 0.82 \\
2240 & 35 & 12 & 0.70 & Lattice & 8 & 15440 & 11520 & 0.75 \\
2560 & 40 & 12 & 0.80 & Lattice & 8 & 12160 &  7680 & 0.63 \\
2880 & 45 & 12 & 0.90 & Lattice & 8 &  8880 &  3840 & 0.43 \\
\bottomrule
\end{tabular}
\caption{Generation parameters for set of graphs in Figure ~\ref{fig.latticevscliques} - Right.}\label{faab-table-lattice}
\end{center}
\end{table}
\end{widetext}

\end{document}